\documentclass[a4paper,11pt]{article}
\usepackage{verbatim}
\usepackage{amsmath,amsthm,amsfonts}
\usepackage{graphicx}
\usepackage{isolatin1}
\usepackage{amssymb}
\usepackage{color}
\usepackage{fancybox}
\usepackage{mathrsfs}	
\usepackage[mathcal]{euscript}
\usepackage{caption}



\newcommand{\e}{{\bf e}}
\newcommand{\dd}{{\sf d}}
\newcommand{\ig}{{\sf IG}}
\newcommand{\N}{\mathbf{N}}
\newcommand{\R}{\mathbf{R}}

\newcommand{\kk}{\kappa }

\newcommand{\tg}{{\sf TG}}
\newcommand{\D}{\mathscr{D}}
\newcommand{\Dd}{{\cal D}}
\newcommand{\A}{\mathcal{A}}
\newcommand{\T}{\mathcal{T}}
\newcommand{\TT}{\mathbf{T}}
\newcommand{\M}{\mathcal{M}}

\newcommand{\ie}{{\it i.e. }}
\newcommand{\cl}[1]{c\hspace{-0.5pt}\ell\!\left(#1\right)}
\newcommand{\sgn}{{\rm sign}\,}

\newcommand{\NN}[1]{\{1\cdots #1\}}
\newcommand{\Nm}[1]{\{0\cdots #1-1\}}

\newtheorem{prop}{Proposition}

\newtheorem{lemma}{Lemma}

\newtheorem{theo}{Theorem}
\newtheorem*{theo2}{Theorem 2}

\begin{document}

\title{Periodic solutions of piecewise affine gene network models:\\
the case of a negative feedback loop} \author{Etienne Farcot, Jean-Luc Gouz\'e}


\maketitle


\abstract{\sl In this paper the existence and unicity of a stable periodic orbit is proven, for a class of
piecewise affine differential equations in dimension 3 or more, provided their interaction structure is a
negative feedback loop. It is also shown that the same systems converge toward a unique stable equilibrium
point in dimension 2. This extends a theorem of Snoussi, which showed the existence of these orbits only.
The considered class of equations is usually studied as a model of gene regulatory networks. It is not
assumed that all decay rates are identical, which is biologically irrelevant, but has been done in the vast
majority of previous studies. Our work relies on classical results about fixed points of monotone, concave
operators acting on positive variables. Moreover, the used techniques are very likely to apply in more
general contexts, opening directions for future work.}

\section{Preliminaries}\label{sec-intro} 
\subsection{Previous works} 
Among recent biological concepts, {\em gene regulatory networks} is a particulary intriguing one. Coarsely
said, it refers to a set of genes coding for proteins being able to activate or inhibit the expression of other
genes within the same set. Since such systems are likely to involve large numbers of genes, and since moreover
gene expression is known to be a non-linear mechanism, mathematical models are, though a required tool, facing
a major challenge. This fact has impelled a large amount of work, giving birth to numerous models,
see~\cite{dejonglong} for a review. Among different types of models, a special class of piecewise linear
differential equations have been proposed in the early 1970's~\cite{glasstopkin}. This class has the advantage
of being mathematically much more tractable than more classical, smooth non-linear equations, nonetheless
offering a comparable variety of behaviours.\\
Since its introduction, the class of piecewise-linear models has thus been studied in various fashions.
Periodic trajectories, or limit cycles, of these systems have been especially investigated. The first results
concerned a particular case, characterized by the equality of all decay rates of the system. Under this assumption,
trajectories are straight lines in regions where the system is affine, and the whole analysis becomes much simpler.
Notably, an explicit Poincaré map is possibly derived, which is a fractional linear map. It is then possible to
reduce the problem of existence and stability of periodic orbits to an eigenanalysis. This method was introduced
in~\cite{glasspastern1,glasspastern2}, and subsequently improved by different authors
\cite{periodsol,edwards,farcot}.\\
 To our knowledge, ony one study did not require this uniform decay rates assumption, but concerned on the other
hand systems whose interaction structure consists in a single, negative loop~\cite{snoussi}. Biological examples,
such as the {\em repressilator}~\cite{letnat1} make this type of interaction structure relevant. For such systems,
the existence, but not unicity, of a stable limit cycle in dimension $\geqslant 3$, and of a stable focus in
dimension $2$ is proven in~\cite{snoussi}. This proof relies on Brouwer's fixed point theorem, and is not entirely
self contained.\\
Our main result is a complete proof of the existence and unicity of the mentioned attractors in feedback loop
systems. Moreover, the mathematical tools involved to prove this result seem to have been ignored in the context
of piecewise linear gene network models, and offer a very promising framework.\\
More precisely, the result on which we mainly rely is a fixed point theorem for monotone, concave operators, where
monotonicity and concavity are defined with respect to a partial order. There exists actually a well-developped
corpus about operators satisfying such properties, especially for systems with positive variables. These results
have the advantage of being generic, and the hypotheses on which they rely are on the other hand 
naturally satisfied in the context of piecewise affine models of gene networks. In order to give a more precise
foretaste of the rest of this paper let us state now the main result. To make it understandable, let us just say
that $\cal C$ is a periodic sequence of regions, crossed by the trajectories of any piecewise affine model of
gene networks consisting in a negative feedback loop. Also, ${\cal Z}^1$ is a particular subset of one region, or
'zone', which is a well defined Poincaré section, and $\TT$ is a first return map on ${\cal Z}^1$.
\begin{theo2}
Let $\TT : {\cal Z}^1 \to {\cal Z}^1$ be the Poincaré map of $\cal C$.
\begin{itemize}
\item If $n=2$, then $\,\forall x \in {\cal Z}^1,\; \TT^m x\to 0$ when $m\to\infty$.
\item If $n>2$, then there exists a unique nonzero fixed point $q=\TT q$.\\[1mm]
Moreover,  $q\in \mathring{\cal Z}^1$ and for every $x\in{\cal Z}^1\setminus\{0\}$, $\,\TT^m x\to q$ as
$m\to\infty$.
\end{itemize}
\end{theo2}

A rapid overview of the tools and references we rely on is given in section~\ref{sec-tools}. Then, the class of
piecewise affine models we focus on is presented in section~\ref{sec-pwaff}. Some general observations about
monotonicity and concavity for this class of systems are made in section~\ref{sec-regions}. These remarks lead to
focus on negative feedback loop systems in section~\ref{sec-negloop}, where the main results are stated and
proven.\\

\subsection{Useful mathematical results}\label{sec-tools}
The main result of this paper strongly rely on one theorem, that we recall here. It is a fixed point theorem of
monotone, concave operators. Both monotonicity and concavity shall be understood with respect to some partial
order, see below for definitions and notations. Operators with both these properties have lead to various results
on existence and unicity of fixed points. Among early studies, Krasnosel'skii has provided important
results~\cite{krasno}. Here, we use a more recent presentation of such results, due to Smith~\cite{smith}, whose
formulation is well adapted to our purpose and context. Namely, we use theorem 2.2 of~\cite{smith}.\\
Before stating the latter, let us introduce some notations (which are similar to those
in~\cite{smith}). We denote $x<y$ and $x\leqslant y$ if these inequalities hold for each coordinate
(resp. entry) of vectors (resp. matrices) $x$ and $y$. Then, we denote $x\lneq y$ if $x\leqslant y$ and
$x \neq y$. For $x\leqslant y$, $[x,y]=\{z\,|\, x\leqslant z\leqslant y\}$, and $(x,y) = \{ z \,|\, x <
z< y\}$. For any set $A$, $\mathring{A}$ denotes the interior of $A$, and $\cl{A}$ its closure. Now, we
can state the following, 
\begin{theo}\label{thm-smith} Let $p\in\mathring{\R}^n_+$, and $T:[0,p] \to [0,p]$
continuous, $C^1$ in $(0,p)$.\\ Suppose
$DT(0)=\displaystyle\lim_{\substack{x\to 0\\ x>0}} DT(x)$ exists.
Assume: 
\begin{itemize} 
	\item[] $\qquad$(M) $\qquad DT(x) > 0$ if $\;x>0$, $x<p$. 
	\item[] $\qquad$(C) $\qquad DT(y) \lneq DT(x)$ if $\;0 < x < y < p$. 
\end{itemize} 
Assume also $Tp < p$.\\ 
Suppose that T0 = 0, and define $\lambda =\rho(DT(0))$, the spectral radius of $DT(0)$. Then,  
\begin{itemize} 
\item[]$\lambda \leqslant 1\implies \forall x\in[0,p]$, $T^n x\to 0$ when $n\to\infty$. 
\item[]$\lambda > 1\implies$ There exists a unique nonzero fixed point $q=Tq$. Moreover, 
$q\in (0,p)$ \\
\phantom{$\lambda > 1\implies$ }and for every $x\in[0,p]\setminus\{0\}$, $T^n x\to q$ as $n\to\infty$. 
\end{itemize} 
\end{theo}

The case $T0\ne 0$ leads to either a unique fixed point, or to diverging orbits. However, this case will not
appear in our context, and thus it is not detailed in the theorem above.\\
One may remark now that in the case when $T$ is twice differentiable, the concavity condition $(C)$
admits a simple sufficient condition.  The map $T_i:[0,p]\to\R_+$ denotes the $i$th coordinate function
of $\, T:[0,p]\to[0,p]$.
\begin{prop}\label{prop-D2C}
Suppose that for all $\;i,j,k\in\NN{n}$, and for all $0< x <p$,
\[
\frac{\partial^2 T_i}{\partial x_k\partial x_j}(x) \leqslant 0,
\]
and for all $i,j$ there exists a $k$ such that the inequality is strict.\\ 
Then $T$ satisfies condition $(C)$ of theorem~\ref{thm-smith}.
\end{prop}
Actually, it is clear that under this condition each term $\frac{\partial T_i}{\partial x_j}$ of $DT$ is a
decreasing function of each coordinate $x_k$. It is moreover strictly decreasing in at least one of these
coordinates, and $(C)$ thus follows. Observe by the way that the notion of concavity (w.r.t a partial order) we
deal with here is weakened by the fact that it concerns only ordered pairs $(x,y)$ of variables.\\
Remarkably, theorem~\ref{thm-smith} relies on Perron-Frobenius theorem, which will be used at some point in the
proof of our main result. It usually says that any matrix with positive entries admits a positive, simple
eigenvalue, associated to a positive eigenvector. Here, we will only need the following corollary: if $A$ is a 
real $n\times n$ matrix such that $A>0$, then $A$ admits a positive real eigenvalue.\\

\section{Piecewise affine models}\label{sec-pwaff}
\subsection{Formulation}\label{sec-prelim}
The general form of the piecewise affine models we consider here may be written as:
\begin{equation}\label{eq-genenet}
\frac{dx}{dt} = \kk(x) - \Gamma x
\end{equation}
The variables $(x_1\dots x_n)$ represent concentrations in proteins or mRNA produced from $n$
interacting genes. Since gene transcriptional regulation is widely supposed to follow a steep
sigmoid law, it has been suggested that idealized, discontinuous switches may be used instead
to model these complex systems~\cite{glasstopkin}. Accordingly, ${\sf
s}^+(\cdot\,,\theta):\R\to \{0,1\}$ denote the increasing step function, or Heaviside
function: 
\[
\left\{\begin{array}{lcl}
{\sf s}^+(x,\theta) & = & 0\quad\text{if } x<\theta,\\
{\sf s}^+(x,\theta) & = & 1\quad\text{if } x\geqslant\theta,
\end{array}\right.
\]
which represent an effect of activation. Also, ${\sf s}^-(x,\theta)=1-{\sf s}^+(x,\theta)$, is its
decreasing version, and represents inhibition.\\
Then, $\kk:\R_+^n\to\R^n_+$ is a piecewise constant production term that can be expressed in terms of step
functions ${\sf s}^\pm(x_i,\theta_{i})$. $\Gamma\in\R_+^{n\times n}$ is a diagonal matrix whose diagonal
entries $\Gamma_{ii}=\gamma_i$, are degradation rates of variables in the system.\\ 
Since each variable $x_i$ is a concentration (of mRNA or of protein), it ranges in some interval of
nonnegative values denoted $[0,\textsf{max}_i]$. When this concentration reaches a threshold value, some
other gene in the network, say gene number $j$, is suddenly produced with a different production rate~:
the value of $\kk_j$ changes. For each $i\in\NN{n}$ there is thus a finite set of threshold values, which
serve as parameter of step functions~:
\begin{equation}\label{eq-thresh}
\Theta_i=\{\theta_{i}^1,\dots,\theta_{i}^{q_i-1}\},
\end{equation}
where the thresholds are ordered: $0<\theta_{i}^1<\dots<\theta_{i}^{q_i-1}<\textsf{max}_i$.  The extreme values $0$
and $\textsf{max}_i$ are not thresholds, since they bound the values of $x_i$, and thus may not be crossed.
However, a convential notation will be : $\theta_{i}^0=0$, and $\theta_i^{q_i}=\mathsf{max}_i$.\\ 
Now, at a time $t$ such that $x_i(t)\in\Theta_i$, there is some $j\in\NN{n}$ such that $\kk_j(x(t^+))\ne
\kk_j(x(t^-))$. It follows that each axis of the state space will be usefully partitionned into open
segments between thresholds. Since the extreme values will not be crossed by the flow (see later), the
first and last segments include one of their endpoints~:
\begin{equation}\label{eq-Di}
{\Dd}_i\in\left\{[\theta_i^0,\,\theta_i^1),\;(\theta_i^{q_i-1},\,\theta_i^{q_i}]\right\}\cup
\left\{(\theta_i^j,\,\theta_i^{j+1})\,|\,j\in\NN{q_i-2}\right\} \cup\Theta_i
\end{equation}
Each product ${\Dd}=\prod_{i=1}^n{\Dd}_i$ defines a rectangular {\it domain}, whose dimension
is the number of ${\Dd}_i$ that are not singletons. When $\dim{\Dd}=n$, one usually says that
it is a {\em regulatory domain}, or {\em regular domain}, and those domain with lower
dimension are called {\em switching domains}~\cite{inria}. We use the notation $\D$ to represent the set
of all domains of the form above. Then, $\D_r$ will denote the set of all regulatory domains,
and $\D_s$ the set of all switching domains.\\ 
A convenient way to represent regular domains makes use of a discrete map, whose range we write as: 
\begin{equation}\label{eq-A}
\A = \prod_{i=1}^n\Nm{q_i}.
\end{equation}
The map in question is then defined as $\dd:\D_r\to\A$, and sends each regular domain to the superscript
of its ``lower-left'' corner: $\dd\left(\prod_i (\theta_i^{a_i-1},\,\theta_i^{a_i})\right)=(a_1-1\dots
a_n-1)$. In the following, we will often identify regular domains and their image in $\A$, talking for
example about some 'domain $a$'.\\ 
The dynamics on regular domains, called here {\it regular dynamics} can be defined quite simply, due to the
simple expression of the flow in each $\Dd\in\D_r$. On sets of $\D_s$ on the other hand, the flow is in
general not uniquely defined. It is anyway possible to define solutions in a rigorous way, yielding what
will be mentioned as the {\it singular dynamics}, at the price of considering set-valued solutions. The
latter's definition rests on Filippov's theory of differential equations with a discontinuous right-hand
side, and its formulation in the particular case of equations~(\ref{eq-genenet}) can be found
in~\cite{casey,gouzesari}, among others. Since we will only encounter systems for which the Filippov
solutions lead to single valued trajectories, we shall only describe the regular dynamics with some
detail.\\

\subsection{Regular dynamics}\label{sec-regdyn}
Regulatory domains are of particular importance. They form the main part of state space, and the dynamics
on them can be expressed quite simply. Actually, on such a domain $\Dd$, the production rate $\kk$ is
constant, and thus equation~(\ref{eq-genenet}) is affine. Its solution is explicitly known, for each
coordinate $i$~:
\begin{equation}\label{eq-flow}
\varphi_i(x,t)=x_i(t) = \frac{\kk_i}{\gamma_i} - e^{-\gamma_i t}\left(x_i -
\frac{\kk_i}{\gamma_i}\right),
\end{equation}
and is valid for all $t\in \R_+$ such that $x(t)\in{\Dd}$. It follows immediately that  
\[
\phi({\Dd}) =
\left(\phi_1\cdots\phi_n\right)=\left(\frac{\kk_1}{\gamma_1}\cdots\frac{\kk_n}{\gamma_n}\right)
\]
is an attractive equilibrium point for the flow~(\ref{eq-flow}). Hence, if it lies inside $\Dd$, it is an
actual equilibrium of system~(\ref{eq-genenet}). Otherwise, the flow will reach the boundary
$\partial{\Dd}$ in finite time, unless $\phi({\Dd})$ lies exactly on the  boundary of $\Dd$. However this
situtation is clearly not generic, and we assume in the rest  of this paper it never occurs. At the time
when the flow reaches $\partial{\Dd}$ thus, the value of $\kk$ will change, and that of $\phi$
accordingly. The point $\phi({\Dd})$ is often called \textit{focal point} of the domain $\Dd$. Then, the
continuous flow can be reduced to a discrete-time dynamical system, with a state space supported by the
boundaries of boxes in $\D_r$.\\ 
Hence, the state space of this discrete-time system is part of $\D_s$, which may seem problematic at first
sight. In fact, it is actually not always possible to define discrete-time trajectories on domains of
$\D_s$ having dimension $n-2$ or less. On any $n-1$ domain $\Dd$, on the other hand, such a definition may
always be properly provided, in terms of the flow lines in the two regular domains separated by $\Dd$. If
these flow lines both point towards, or away from $\Dd$ (which is expressed formally using the flow
coordinate which is normal to $\Dd$), the latter is called respectively {\it black wall} or {\it white
wall}. Otherwise, \ie when flow lines both cross $\Dd$ in the same direction, one usually refers to $\Dd$
as a {\it transparent wall}. In the two first cases, the Filippov theory is required, providing set-valued
trajectories on the walls, the so-called sliding modes~\cite{gouzesari}. On transparent walls,
trajectories are simply defined by continuity, from the flow lines on both sides. We shall encounter only
this third kind of wall in the rest of this paper.\\
Note that if $\Dd\in\D_r$ is represented by $a\in\A$,
\ie $\dd(\Dd)=a$, one shall most often denote by $\phi(a)$, or $\phi^a$, the focal point
associated to this domain.\\

Once the flow (\ref{eq-flow}) is given in a box ${\Dd}_a$, it is easy to compute the time and
position at which it intersects the boundary of ${\Dd}_a$, if ever. The possibility for each
facet to be encountered by the flow depends uniquely on the position of the focal point~:
$\{x\,|\,x_i=\theta_{i}^{a_i-1}\}$ (resp. $\{x\,|\,x_i=\theta_{i}^{a_i}\}$) can be crossed if
and only if $\phi_i<\theta_{i}^{a_i-1}$ (resp. $\phi_i>\theta_{i}^{a_i}$). According to this
observation, we denote $I_{out}^+(a) = \{i\in\NN{n} | \,\phi_i >\theta_{i}^{a_i}\}$, and
$I_{out}^-(a) = \{i\in\NN{n} | \,\phi_i <\theta_{i}^{a_i-1}\}$.\\ 
Then, $I_{out}(a)=I_{out}^+(a)\cup I_{out}^- (a)$ is the set of escaping directions of $\Dd_a$.\\
Also, it will be practical to point out the lower and upper thresholds bounding a box, thanks
to the pairs of functions $\theta_i^\pm:\A\to\Theta_i$, $\theta_i^-(a)=\theta_{i}^{a_i-1}$
and  $\theta_i^+(a)=\theta_{i}^{a_i}$. These functions will clearly only be useful notations,
and bring no further information about the system.\\ When it is unambiguous, we will omit the
dependence on $a$ in the sequel.\\ Now, in each direction $i\in I_{out}$ the time at which
$x(t)$ encounters the corresponding hyperplane, for $x\in {\Dd}_a$, can easily be shown to
be:
\begin{equation}\label{eq-taui}
\tau_i(x)=\frac{-1}{\gamma_i}\ln\left(\min\left\{\frac{\phi_i -\theta_{i}^-(a)}{\phi_i -x_i},
\frac{\phi_i -\theta_{i}^+(a)}{\phi_i -x_i}\right\}\right). 
\end{equation}
Taking the minimum 
$\tau(x)=\min_{i\in I_{out}}\tau_i(x)$,
and reinjecting it in equation (\ref{eq-flow}), we get the exiting point of ${\Dd}_a$ when
starting at $x$. Since this process is intended to be repeated along trajectories, $x$ will
generally lie on the boundary of the current box, except for the initial condition, which may
however be chosen on a facet without loss of generality. We then get a \textit{ transition map}
${\T}^a: \partial \Dd_a\rightarrow \partial \Dd_a$~:
\begin{equation}\label{eq-maptrans}
\begin{array}{lcl}
{\T}^ax & = & \varphi\left(x,\tau(x)\right)\\
          & = & \phi + \alpha(x) (x-\phi).
\end{array}
\end{equation}
where $\alpha(x) = \exp(-\tau(x)\Gamma)$. The latter depends on $a$, as seen
from~(\ref{eq-taui}).\\ 
Now, the initial system~(\ref{eq-genenet}) may been reduced to a discrete time dynamical system, as
mentioned earlier. It consists in iterates of a global map $\T$ on a domain ${\rm Dom}_{\,\T}$. As
already mentioned, ${\rm Dom}_{\,\T}$ must be contained in the set of $n-1$ domains of $\D_s$. Since
$\T$ has to be iterated on this domain, all points which reach some $n-2$ dimensional (or less) domain
after applying $\T$ a finite number of times, must be excluded from ${\rm Dom}_{\,\T}$. Furthermore,
whereas $\T$ is readily defined on transparent walls, black and white walls are much less obvious to
deal with. We thus suppose from now on that ${\rm Dom}_{\,\T}$ entirely consists of transparent walls. A
more detailed discussion about the topology of this domain can be found in~\cite{farcot}. A simple criterion to
ensure that all walls are transparent is the absence of autoregulation, in the sense that no production term
$\kk_i$ depends on $x_i$.\\ 
At each point of such a domain, it is possible to consistently apply a unique local map ${\T}^a$.
Actually, each wall is either on the boundary of the whole domain $\bigcup_a\Dd_a$, and thus of a single
regular domain $\Dd_a$, or can be written as an intersection $\partial\Dd_a\cap \partial\Dd_b$ of two
regular domain boundaries. It is clear that on transparent walls there is always exactly one of the two
domains, say $\Dd_a$, such that ${\T}^a$ is not the identity. Applying ${\T}^a$ is then consistent with
the orientation of the flow lines. Yet, we only care here with forward trajectories, since ${\T}^{-1}$
is not properly defined on the full domain ${\rm Dom}_{\,\T}$.\\ 
Suppose now that there is a wall $W\subset{\rm Dom}_{\,\T}$, and a sequence $a^1\dots a^\ell$ of regular domains
such that $({\T}^{a^\ell}\circ{\T}^{a^{\ell-1}}\cdots\circ{\T}^{a^1}) (W)\cap W\ne \varnothing$. Now, it is
tempting to study the dynamics of this iterated map on $W$. In particular, any fixed point of this map 
determines a periodic trajectory of the continuous time system~(\ref{eq-genenet}).

\section{Regions of fixed monotonicity and concavity}\label{sec-regions}
In order to apply theorem~\ref{thm-smith} to an iterate of the transition map, as defined in
previous section, it is necessary to check whether it is monotone and concave, in the sense of
properties $(M)$ and $(C)$. In this section it is shown that these conditions cannot be satisfied by the
transition map on a whole wall, and that some refinement should be done. To this aim, one should compute
the jacobian of a composite map, of the form: $D({\T}^{a^\ell}\circ{\T}^{a^{\ell-1}}\cdots\circ{\T}^{a^1})$.
This requires the computation of an arbitrary $D{\T^{a^i}}$.\\
Let $W$ be an arbitrary wall. One has seen in previous section that ${\T}={\T}^a$, for some uniquely
defined $a$. Let moreover $x\in W$ such that there exists a unique $s\in I_{out}(a)$ such that
$\tau(x)=\tau_s(x)$, and denote $W'\subset\{x\,|\,x_s=\theta_s\}$ the wall such that $\T x\in W'$. Then
there exists an open neighbourhood of $x$ in which the following holds:
\begin{equation}\label{eq-Tia}
{\T}_ix = \phi_i^a + (x_i - \phi^a_i)\, \alpha_i(x),
\end{equation}
where ${\T}_i$ is the $i$th coordinate function of $\T$. To abbreviate computations, we have denoted
$\alpha_i$ the $i$th diagonal entry of the diagonal matrix $\alpha(x)$ defined in previous section:
\begin{equation}\label{eq-alphai}
\alpha_i(x) = \left(\frac{\phi^a_s - \theta_s}{\phi^a_s - x_s}\right)^ {\frac{\gamma_i}{\gamma_s}}.
\end{equation}
Then it is rather straightforward to compute the partial derivatives at a point $x$:
\begin{equation}\label{eq-dTidxj}
\frac{\partial\T_i}{\partial x_j} =\left\{
\begin{array}{ll}
\displaystyle\left(\alpha_s(x)\right)^{\frac{\gamma_i}{\gamma_s}} & \text{if }\;  i=j\\[2mm]
\displaystyle-\frac{\gamma_i}{\gamma_s}\frac{\phi_i^a - x_i}{\phi_s^a - x_s}
\left(\alpha_s(x)\right)^{\frac{\gamma_i}{\gamma_s}} & \text{if }\;  j=s\\[2mm]
0 & \text{otherwise}.
\end{array}\right.
\end{equation}
It appears from~(\ref{eq-dTidxj}) that the partial derivative are of constant sign if, and only if,
$\sgn(\phi^a_i-x_i)$ is fixed. Actually, all degradation rates $\gamma_i$ are positive, and
$\alpha_s(x)=\exp(-\gamma_s \tau(x))\in (0,1)$ since $\tau(x)$ is nonnegative.\\

Now make two observations on the sign pattern of $\phi^a-x$ in general. The first one concerns its invariance
with respect to time evolution:
\begin{prop}\label{rem-sgnpreserv}
From the expression~(\ref{eq-flow}), one can observe that the quantities $\sgn(\phi^a_i-x_i)$ are
preserved by the flow, in the sense that $\sgn(\phi^a_i-x_i)=\sgn(\phi^a_i-\varphi_i(x,t))$ for any
$t\geqslant 0$.\\
In particular, since the wall $W$ chosen at the beginning of this section is included in a hyperplane of
the form $\{x\,|\,x_j=\theta_j\}$, the above observation yields
$
\sgn(\phi^a_j - \T_j x)=\sgn(\phi^a_j - \theta_j).
$\\
Similarly, for any $x$ which escapes via the wall $W'\subset\{x\,|\, x_s=\theta_s\}$, one has 
$ 
\sgn(\phi_s^a-x_s)=\sgn(\phi_s^a- \theta_s).
$ \\ 
\end{prop}
A second fact concerns the influence of the different coordinates on $\sgn(\phi^a-x)$.
\begin{prop}\label{rem-phisi}
For any escaping direction $i\in I_{out}(a)$, one has either $x_i<\theta_i^+(a)<\phi_i^a$, or
$\phi_i^a<\theta_i^- (a) < x_i$, and in both cases $\sgn(\phi_i^a-x_i)$ is constant in the whole closure
$\cl{\Dd_a}$ of the domain. If $\theta_i\in\{\theta_i^-(a),\,\theta_i^+(a)\}$ is the escaping threshlod,
this sign equals $\sgn(\phi_i^a- \theta_i)$.\\
For the non escaping directions on the other hand, $\sgn(\phi_i^a-x_i)$ is not constant on $W$, since
$\theta_i^-(a) < \phi_i^a < \theta_i^+(a)$. \\
\end{prop}
Thus, condition $(M)$ cannot be valid on the whole wall $W$, unless all directions are escaping.\\

Since a differential of the form $D({\T}^{a^\ell}\circ{\T}^{a^{\ell-1}}\cdots\circ{\T}^{a^1})$ is a product
of  differentials of local maps $D({\T}^{a^i})$, and since the latter may only be positive in a region where
the vector $\phi^{a^i} - x$ has a fixed sign pattern, it seems relevant to restrict the study to such
regions.\\
Now, condition $(C)$ concerns the variations of the terms in~(\ref{eq-dTidxj}). As shown by
proposition~\ref{prop-D2C}, the second-order derivatives are informative in this case. They can be
expressed as:
\begin{equation}\label{eq-d2Tidxj}
\frac{\partial^2\T_i}{\partial x_m\partial x_j} =\left\{
\begin{array}{ll}
\displaystyle\frac{\gamma_i}{\gamma_s} \frac{\left(\alpha_s(x)\right)^{\frac{\gamma_i}{\gamma_s}}}
{\phi_s-x_s}
& \text{if }\;  i=j,\,m=s\quad\text{or } j=s,\,m=i\\[3mm]
\displaystyle\frac{\gamma_i}{\gamma_s} \left(1+\frac{\gamma_i}{\gamma_s}\right) \frac{\phi_i^a -
x_i}{(\phi_s^a - x_s)^2} \left( \alpha_s(x) \right) ^{\frac{\gamma_i}{\gamma_s}} & \text{if }\; 
m=j=s\\[3mm]
0 & \text{otherwise}.
\end{array}\right.
\end{equation}
Here again, these derivatives are of fixed sign only if $\phi^a-x$ has a fixed sign pattern. Hence, both first
order and second order derivatives are of constant sign in the same regions. One is thus naturally led to
consider rectangular regions partitioning the wall $W^i$, defined by the sign of $\phi^a-x$, or in other words by
the position (in terms of the partial order) of points with respect to the focal point.\\

It will be shown in the following sections that in the special case of negative feedback loop systems, only
one such region contains the long-term dynamics, on each wall of a periodic sequence of domains in state
space. Before proving this, let us discuss a slightly more general case than feedback loop systems. Our hope
here is that this discussion give some intuition of the basic facts that will serve afterwards. It also
suggests that tools from positive operators theory might be efficient in a broader context than the sole
feedback loop systems.\\
Namely, suppose given a periodic sequence of domains $a^1, \dots a^\ell, a^{\ell+1}=a^1$, such that there
exists a unique escaping direction for each box $a^i$. This case is actually the simplest one when studying
periodic orbits, since if there are several escaping directions for one box in the sequence, then some
trajectories escape the cycle at this box. The set of points remaining in the periodic sequence $a^1\dots
a^\ell$ is not easily defined in general, though under the uniform decay rates assumption it may be expressed
by a list of affine inequalities~\cite{edwards,farcot}.\\
Then, each wall reached by successive iterates of the transition map has to be partitioned into $2^{n-2}$
regions in general. Actually, on such a wall points have $n-1$ coordinates which are not constant: in one
direction it equals a threshold value. Among these $n-1$, one exactly is the escaping direction $s_{i-1}$ of
the previous box, $a^{i-1}$. Thus according to proposition~\ref{rem-sgnpreserv}, the following is fixed
\[
\sgn(\phi_{s_{i-1}}^{a^i}-x_{s_{i-1}})=\sgn(\phi_{s_{i-1}}^{a^i}-\theta^{m}_{s_{i-1}}),
\]
for some $m$, and any point $x$ on a trajectory crossing the considered sequence of domains.\\
Thus if one assumes $s_i\ne s_{i-1}$, then $n-2$ directions remain for which two signs are possible. At each
wall, the images of these $2^{n-2}$ region may intersect several regions among the $2^{n-2}$ partitioning the
next wall. Hence, one may refine the partition of the latter, and iterating this process, obtain a partition of a
wall into regions such that all partial derivatives of iterates, of the form $\frac{\partial\T_i^k} {\partial
x_j}$, for $k\in\NN{\ell}$ are of fixed sign, and the second order derivatives as well. This process would lead
to consider a high number of regions in general, each of which may satisfy the hypotheses of~\ref{thm-smith}.
This seems far from trivial in the most general case, but would certainly be fruitful in some more particular
cases.\\
Hence, let us look for such special cases of equations~(\ref{eq-genenet}). For two successive boxes $a^{i-1}$ and
$a^{i}$, one has, as seen from~(\ref{eq-Tia}):
\begin{equation}\label{eq-alfoc}
\phi_j^{a^{i}} - {\T}_jx = \phi_j^{a^i} - \phi_j^{a^{i-1}} + (\phi^{a^{i-1}}_j - x_j)\, \alpha_j(x),
\end{equation}
and if the focal points are `aligned`, in the sense that $\phi_j^{a^i}= \phi_j^{a^{i-1}}$, the sign of the
quantity $\phi_j-x_j$ remains unchanged after applying $\T$. Actually one as seen already that $\alpha_s(x)\in
(0,1)$, and applying $\T$ implies that the focal point must be replaced by that of the following box. This
geometric condition should thus avoid us to refine the partitions of successive walls, as mentioned in the above
discussion, since it imposes that these regions do not overlap from one wall to the following.\\
It may be stated more formally. Recall that $a^{i}$ refers to the $i$th crossed box, and that $s_i$ is its unique
escaping direction. Assume a strong version of the previous alignment of focal points:
\begin{equation}\label{eq-align}
\forall i\in\NN{\ell},\,\forall j\in \NN{n}\setminus\{s_{i+1}\},\quad 
\phi^{a^i}_j = \phi^{a^{i+1}}_j.
\end{equation}
This alignment assumption will allow us to use theorem~\ref{thm-smith} in the region where the
long-term dynamics take place. Furthermore, it will be shown that the geometric property
(\ref{eq-align}) can also be obtained as a consequence of further assumptions on the interaction 
structure of the system: if the latter consists in a single negative feedback loop, it follows that 
there exists a periodic sequence of boxes satisfying~(\ref{eq-align}), as we will see in the next section.\\

\section{Negative feedback loop systems}\label{sec-negloop}
\subsection{Preliminary definitions}\label{sec-negloopdef}
In this section, we focus on systems of the form:
\begin{equation}\label{eq-sysloop}
\left\{\begin{array}{lcll}
\displaystyle\frac{dx_1}{dt} & = & \kk_1^0+\kk_1^1\,{\sf s}^-(x_n,\theta_{n}^1) - \gamma_1 x_1 & \\[2mm]
\displaystyle\frac{dx_i}{dt} & = & \kk_i^0+\kk_i^1\,{\sf s}^+(x_{i-1},\theta_{i-1}^1) - \gamma_i x_i,
&\qquad i=2\dots n.
\end{array}\right.
\end{equation}
In such systems, each variable $x_i$ activates the production of the next variable, $x_{i+1}$, except
$x_n$ which inhibits $x_1$. Hence, their interaction structure is that of a negative feedback loop. Up
to a translation, they are of the form dealt with by Snoussi in~\cite{snoussi}. In the case
$n=3$ one finds the well-known {\em repressilator}~\cite{letnat1}. Remark now that any production term $\kk_i$
is independent on the variable $x_i$, and thus all walls must be transparent.\\
Without introducing  unnecessary formalism, let us only say that we call {\em interaction graph}, $\ig$
is an oriented,  labeled graph, whose vertices represent variables of the system, and edges represent
interactions between them. For example,  $ x_i \longrightarrow^- x_j $ if an increase
of $x_i$ may lead to a decrease of $\kk_j$ at least at some point $x$ in state space. Then, a negative
loop is a cycle in $\ig$ with an odd number of negative edges. It is easily shown that any system whose
interaction graph consists in a single negative loop (involving all variables) is equivalent
to~(\ref{eq-sysloop}), up to a symmetry in state space.\\ 
The particular form of~(\ref{eq-sysloop}) implies that there is only one threshold $\theta_i^1$ in each
direction $i$, and $\theta_i^0$, $\theta_i^2$ represent the bounds of the range of $x_i$, in accordance
with~(\ref{eq-thresh}).\\
In the following we always assume that
\begin{equation}\label{eq-focconstr}
\forall i\in\NN{n} \qquad \kk_i^0+\kk_i^1 > \gamma_i\,\theta_i^1\quad \text{and}\quad
\kk_i^0 < \gamma_i\,\theta_i^1
\end{equation}
This implies that higher values of focal points are above thresholds, and lower values are below thresholds.
Otherwise, all trajectories would converge to a unique fixed point, see~\cite{snoussi} lemma~1.\\

Before entering into more detail, let us operate a few simplifying operations. First, we chose to set
$w_j=x_j-\theta_j^1$. Then, 
\[
\dot w_j = \kk_j^0+\kk_j^1{\sf h}^\pm(w_{j-1},0)-\gamma_j\theta_j^1 - \gamma_jw_j,
\]
and thus all treshold values are zero. Accordingly, all regular domains are now identified with (bounded
rectangular regions of) the $2^n$ orthants of $\R^n$. They are mapped as explained in
section~\ref{sec-prelim}, onto the discrete set $\A=\{0,1\}^n$. The corresponding discrete mapping
$\dd:\D_r\to\A$ sends negative coordinates to $0$, and positive ones to $1$.\\
The bounds of the whole domain are now denoted 
\[
\theta_i^-=\theta_i^0-\theta_i^1=-\theta_i^1\quad\text{and}\qquad\theta_i^+=\theta_i^2-\theta_i^1.
\] 
Observe that variables may now be negative, and do represent concentrations only up to a translation.\\
The new dynamical system behaves identically to~(\ref{eq-sysloop}), once thresholds have been set to
zero, and the following renamings have been done:
\[
\kk_i^0\leftarrow \kk_i^0 - \gamma_i\theta_i^1,\qquad\text{and}\qquad\phi_i \leftarrow \phi_i-\theta_i^1.
\]
Then, for each $i\in\NN{n}$, the focal point coordinate $\phi_i$ may only take two values. We abbreviate
them as follows:
\begin{equation}\label{eq-phipm}
\phi_i^- = \frac{\kk_i^0}{\gamma_i}\qquad\text{and}\qquad \phi_i^+ = \frac{\kk_i^0+\kk_i^1}{\gamma_i}.
\end{equation}
One shall also note $\kk_i^\pm=\gamma_i\,\phi_i^\pm$.\\
Remark that $\phi_i^-$ is negative and $\phi_i^+$ positive (and similarly for $\kk_i^\pm$).  From this
observation, one deduces that the focal points are entirely determined by the sign of their coordinates. In
other words, the coordinates of any focal point are known exactly from the regular domain it belongs to. More
precisely, if a focal points belongs to a domain $a\in\A$, the sign of $\phi_i$ is $2a_i -1\in\{\pm 1\}$, for
each $i\in\NN{n}$.\\

Now it is time to describe the dynamics of~(\ref{eq-sysloop}). In brief, we now focus on systems of the
form below, into which any  system~(\ref{eq-sysloop}) can be transformed:
\begin{equation}\label{eq-sysloopp}
\left\{\begin{array}{lcll}
\displaystyle\frac{dx_1}{dt} & = & \kk_1^-+(\kk_1^+-\kk_1^-)\,{\sf s}^-(x_n,0) - \gamma_1 x_1 & \\[2mm]
\displaystyle\frac{dx_i}{dt} & = & \kk_i^-+(\kk_i^+-\kk_i^-)\,{\sf s}^+(x_{i-1},0) - \gamma_i x_i,
&\qquad i=2\dots n.
\end{array}\right.
\end{equation}

\begin{center}
\begin{figure}
\begin{center}
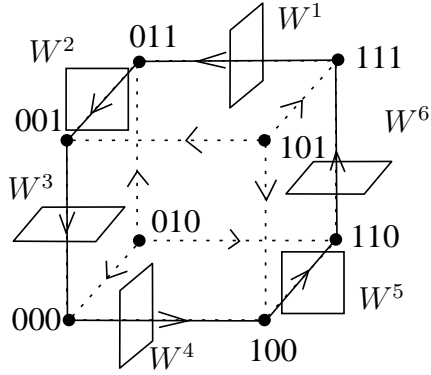
\caption{\label{fig-tg3d}The transition graph of a feedback loop system of the form~(\ref{eq-sysloopp})
with three variables. The cycle $\cal C$ is represented with plain lines, and the walls $W^i$ have been
schematically depicted as well.}
\end{center}
\end{figure}
\end{center}

A first, coarse-grained description can be achieved in terms of discrete transitions between qualitative
states in $\A$. This can be conveniently explained with the aid of the {\em state transition graph}, or
simply {\em transition graph}, whose vertices are elements of $\A$, and edges $(a,b)$ exist if and only if
there exists some continuous trajectory crossing successively the corresponding regular domains in state
space, $\Dd_a$ and $\Dd_b$. This graph will be denoted $\tg$.\\
It is easily deduced from the properties presented in section~\ref{sec-regdyn} that $\tg$ is entirely
determined by the position of focal points. More precisely, $(a,b)\in{\cal E}$ if and only if
$b=a=\phi(a)$, or
\[
b\in\{a+\e_i\,|\,i\text{ such that } a_i=0\text{ and }\phi_i(a) > 0\}\cup 
\{a-\e_i\,|\,i\text{ such that } a_i=1\text{ and }\phi_i(a) < 0\}.
\]
Observe that $a$ and $b$ may differ at one coordinate at most, since $\tg$ represents the regular dynamics
only.\\
From~(\ref{eq-sysloopp}) and~(\ref{eq-focconstr}), one can deduce the presence of the following cycle of
length $2n$ in $\tg$:  
\[
{\cal C}= \begin{array}{ccccccccc}  1\cdots 11 &\longrightarrow & 01\cdots 1&\longrightarrow & \cdots & 
\longrightarrow &  0\cdots 011 & \longrightarrow & 0\cdots 01 \\
\Big\uparrow & & & & & & & & \Big\downarrow \\
  1\cdots 10 & \longleftarrow  & \cdots & \longleftarrow & 110\cdots 0 & \longleftarrow &  10\cdots 0 & 
  \longleftarrow &   00\cdots 0  
\end{array}
\]
It can be shown that there is no fixed state in $\tg$, and that $\cal C$ is the only cycle without
escaping edge. Then, we denote
\begin{equation}\label{eq-C} 
\begin{array}{c|c|c|c|c|c|c|c|c}
a^1 & a^2 & a^3 & \dots & a^n & a^{n+1} & \dots & a^{2n-1} & a^{2n} \\
\hline
1\cdots 11 & 01\cdots 1 &  001\cdots 1 & \dots & 0\cdots 01  &  0\cdots 0 & \dots &  1\cdots 100 & 1\cdots 10
^{\phantom{\big(}}
\end{array}
\end{equation}
Furthermore, let us write $\phi^i=\phi(a^i)$, the focal point of domain $a^i$, and $W^i$ the
wall between boxes $a^{i}$ and $a^{i+1}$, \ie it verifies $\cl{W^i}= \cl{\Dd_{a^{i}}}\cap
\cl{\Dd_{a^{i+1}}}$.\\
In the following, the superscript in $a^i$ must be understood periodically, so that for instance
$a^{2n+1}$ will mean $a^1$. It appears that each pair of successive states $a^i,\,a^{i+1}$ in this cycle
differ only at one coordinate, which we denote $s_i$. If moreover we denote $s_i^\pm$ where $\pm$ is the
sign of $a^{i+1}_{s_i}-a^i_{s_i}$, which is also the sign of $\phi^i_{s_i}$, one obtains
\begin{equation}\label{eq-sipm}
\begin{array}{l|cccccc}
i         &  1 & \cdots & n & n+1 & \cdots & 2n \\
\hline
s_{i}^\pm & 1^- & \cdots & n^- & 1^+ & \cdots & n^{+\phantom{)^i}}
\end{array}
\end{equation}
Hence $s_i$ is the only escaping direction of box $a^i$, a fact that can also be written as
$I_{out}(a^i) = \{s_i\}$. As seen in section~\ref{sec-regdyn}, this implies that all trajectories in a
domain ${a^i}$ leave it in finite time, via the wall $ W^{i}\subset\{x\,|\,x_{s_i}=0\}$, and
enter domain ${a^{i+1}}$. It follows that no trajectory can escape the cycle, \ie ${\cal C}$ is an
invariant region. This cycle is represented on figure~\ref{fig-tg3d} for $n=3$.\\
A rapid inspection shows that the walls can be described in a quite explicit manner:
\begin{equation}\label{eq-Wi}
W^i=
\left\{\begin{array}{ll}
	\displaystyle \prod_{j<s_i}[\theta_j^-,\,0)\times\{0\}\times\prod_{j>s_i}(0,\,\theta_j^+] & 
	\text{for}\;i\in\NN{n},\\[4mm]
	\displaystyle \prod_{j<s_i}(0,\,\theta_j^+]\times\{0\}\times\prod_{j>s_i}[\theta_j^-,\,0) & 
	\text{for}\;i\in\{n+1\cdots 2n\}.
\end{array}\right.
\end{equation}

From the remark under equation~(\ref{eq-phipm}) it is not difficult to deduce that each pair of focal
points $\phi^i$, $\phi^{i+1}$ differ only at coordinate $s_{i+1}$. Actually, they verify
$\phi^i\in\Dd_{a^{i+1}}$, and one has seen that $a^{i+1}$ and $a^{i+2}$ only differ at coordinate
$s_{i+1}$. Hence, the cycle $\cal C$ satisfies the alignment condition~(\ref{eq-align}) seen in previous 
section. We can state it with our current notations:
\begin{equation}\label{eq-alignloop}
\forall i\in\NN{2n},\,\forall j\in \NN{n}\setminus\{s_{i+1}\},\quad \phi^{i}_j = \phi^{i+1}_j.
\end{equation}
Remark also that by construction one has for the remaining coordinate:
\begin{equation}\label{eq-alignloopbis}
\forall i\in\NN{2n},\quad \phi^{i}_{s_{i+1}} = -\phi^{i+1}_{s_{i+1}}.
\end{equation}

Since our aim is to study limit cycles arising from~(\ref{eq-sysloop}), with the help of
theorem~\ref{thm-smith}, it seems natural to consider the following map:
\[
\T^{a^1}\T^{a^{2n}}\T^{a^{2n-1}}\cdots \T^{a^2}: W^1 \longrightarrow W^1
\]
Although each $\T^a$ is defined on the whole boundary of $\Dd_a$, see~(\ref{eq-maptrans}), one 
considers here the restrictions to walls $W^{i-1}$ only, which we write explicitly:
\begin{equation}\label{eq-Tai}
\begin{array}{llcl}
{\T}^{a^i}: & W^{i-1} & \longrightarrow & W^{i} \\
            & x   & \longmapsto     & 
\Big(\, \phi^i_j + (x_j - \phi^i_j)\alpha_j(x) \,\Big)_{j=1\dots n},
\end{array}
\end{equation}
where, as in~(\ref{eq-alphai}), 
\[
\alpha_j(x)=\alpha_j(x_{s_i})=\left(\frac {\phi^i_{s_i}} {\phi^i_{s_i} - x_{s_i}}
 \right)^{\frac{\gamma_j}{\gamma_{s_i}}}.
\]
Remark here that the expression of $\alpha_j$ depends on the number $i$ of the region under
consideration. However, this region will always be clear from the context, and we thus leave this
dependence hidden, avoiding notations such as $\alpha_j^i(x)$.\\

\subsection{Restriction of domains}\label{sec-restricdom}
\subsubsection{Projection and extension to the closure of walls}

\begin{figure}
\begin{center}
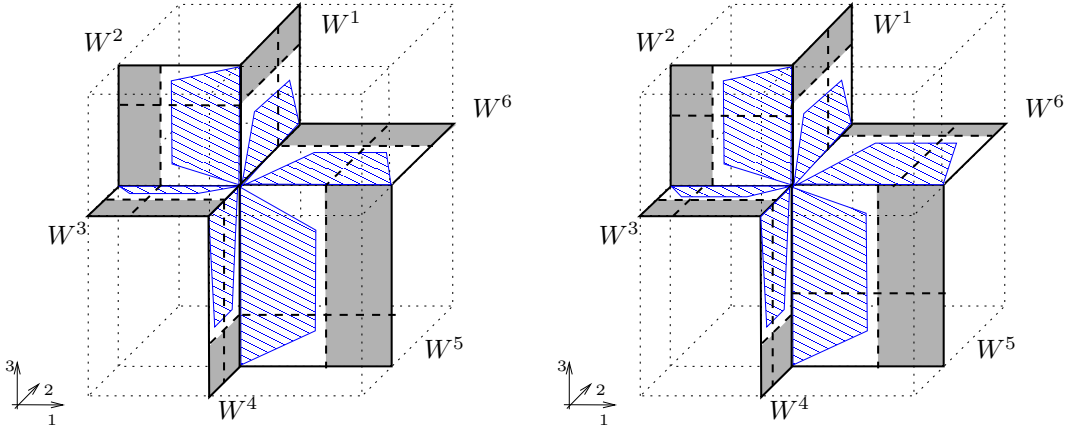
\input{Figures/sympetalesnonalign.pstex_t}
\caption{\label{fig-firstcut}The same walls as in figure~\ref{fig-tg3d}, this time in phase space. Dotted lines
represent thresholds. The dashed lines on a wall $W^{i}$ represent lines where $x_j=\phi_j^i$. The alignment
condition~(\ref{eq-alignloop}) is fulfilled on the left, and not on the right hand side. The right-hand side
figure is thus only presented as a counterexample to~(\ref{eq-alignloop}). The blue region on each such wall
represents the image $\T^{a^i}\left(W^{i-1}\right)$. They are shown as polyhedras for ease of observation, but
they have in general a piecewise smooth boundary, and not a piecewise linear one (unless all decay rates are
equal). The shaded regions are not reachable, as an illustration of eq.~(\ref{eq-sgnphisi-1}).}
\end{center}
\end{figure}

To fit the hypotheses of theorem~\ref{thm-smith}, one must consider a closed domain of the form
$[0,p],\,p\in\R_+^n$, with nonempty interior, and since the walls are $n-1$ dimensional, their interior
is empty.\\
Therefore, one may first project $W^1$ to $\R^{n-1}$, and drop out the first coordinate,
which is zero. Similarly, there is no loss in suppressing the $s_{i}$th coordinate of points in $W^i$,
since it equals zero by construction.\\
Second, each $\T^{a^i}$ has an extension to the closure $\cl{W^{i-1}}$, defined by continuity of the
expression~(\ref{eq-Tai}). To compose these extensions, one has to check that each of them maps
$\cl{W^{i-1}}$ to $\cl{W^{i}}$. Since the walls themselves are correctly mapped, and are products of one
singleton $\{0\}$ and of segments of the form $(0,\theta_j^+]$ or  $[\theta_j^-,0)$, it suffices to check
that the points of $\cl{W^{i-1}}$ having at least one coordinate equal to $0$ are mapped to
$\cl{W^{i}}$.\\
So, let $x\in\cl{W^{i-1}}$ such that $x_j=0$. Since $\cl{W^{i-1}}\subset\{x\,|\,x_{s_{i-1}}=0\}$ it is
relevant to consider only $j\ne s_{i-1}$. Then, $\T^{a^i}x\in\cl{W^{i}}$ if, and only if, each coordinate
$\T^{a^i}_kx$ belongs to the projection in direction $k$ of $\cl{W^{i}}$, which is either
$[0,\theta_k^+]$ or $[\theta_k^-,0]$. Let us call this segment ${\cal I}_k$ to simplify the exposition.
The  explicit formula given in~(\ref{eq-Tai}) shows that the coordinate map $\T^{a^i}_j$ depends only on
variables $x_j$ and $x_{s_i}$. Hence, if $j\ne s_i$, one only has to check whether $\T^{a^i}_jx\in {\cal
I}_j$, while if $j=s_i$ all coordinates must be checked. These two cases give
\begin{itemize}
\item If $j\ne s_i$, $\T^{a^i}_jx=(1-\alpha_j(x))\,\phi_j^i$. From the inequalities between $x_j$, 
$\phi_j^i$ and $\theta_j=0$, it is easily checked that $\alpha_j(x)\in (0,1]$. Moreover, for $j\not\in
\{s_i\}=I_{out}(a^i)$, $\phi^i_j$ is in ${\cal I}_j\setminus\{0\}$. Then, $\T^{a^i}_jx\in{\cal
I}_j$ follows easily.
\item If $j=s_{i}$, remark first that $x\in\cl{W^{i-1}}\cap\cl{W^{i}}$. Then $\T^{a^i}_{s_{i}}x =
0\in{\cal I}_j$. Moreover, for any $k\in\NN{n}$, $\alpha_{s_{k}}(x)=1$, so that $\T^{a^i}_kx=x_k$, that
is to say $x$ is a fixed point.
\end{itemize}
It follows that in any case, $\T^{a^i}x\in\cl{W^{i}}$, as awaited, and we consider now that the maps
$\T^{a^i}$ are defined on the closure of walls.\\
Finally, proposition~\ref{rem-sgnpreserv} implies that the image $\T^{a^i}\left(\cl{W^{i-1}}\right)$ is always
contained in the subset of $\cl{W^{i}}$ whose points satisfy 
\begin{equation}\label{eq-sgnphisi-1}
\sgn(\phi_{s_{i-1}}^i-x_{s_{i-1}}) =\sgn(\phi_{s_{i-1}}^i).
\end{equation}
This restriction is illustrated on figure~\ref{fig-firstcut}. Since our interest is in the iterates of
those maps, we should restrict them to such subsets.\\
In summary, three simplifications can be done:
\begin{itemize}
\item The $s_{i}$th coordinate of $x\in W^i$ can be suppressed without loss of information
\item The transition maps can be defined on the closure of walls.
\item Only the points verifying~(\ref{eq-sgnphisi-1}) can be reached by $\T^{a^i}$.
\end{itemize}
The notation $\widetilde W^{i}$ will be used to denote the domains obtained after these three
simplifications. More explicitly $\widetilde W^i$ is obtained from~(\ref{eq-Wi}) by suppressing the
$\{0\}$ term, using closed intervals, and replacing moreover $[\theta_{s_{i-1}}^-,0]$ by 
$[\phi_{s_{i-1}}^-,0]$, or $[0,\theta_{s_{i-1}}^+]$ by $[0,\phi_{s_{i-1}}^+]$. For example:
\[
W^1=\{0\}\times \prod_{i=2}^n(0,\,\theta_i^+],\quad\text{so that}\qquad
\widetilde W^1 = \prod_{i=2}^{n-1}[0,\,\theta_i^+]\times[0,\phi_{n}^+].
\]

Each $\widetilde W^i$ is a rectangular region in $\R^{n-1}$ with nonempty interior, still called wall in
the sequel, and the preceding discussion shows that each map $\T^{a^i}$ induces a well defined map 
\[
\widetilde\T^{a^i} : \widetilde W^{i-1} \to \widetilde W^{i}.
\]
Remark that although points in $\widetilde W^{i-1}$ do not have a $s_{i-1}$th coordinate, 
$\widetilde\T^{a^i}_{s_{i-1}}$ is defined using expression~(\ref{eq-Tai}), with $x_{s_{i-1}}=0$. Let us
write it explicitly for later purpose:
\begin{equation}\label{eq-Taisi-1}
\widetilde\T^{a^i}_{s_{i-1}}x = \left(1- \alpha_{s_{i-1}}(x)\right)\phi^i_{s_{i-1}}
\end{equation}
Now, we define a map
\begin{equation}\label{eq-TT}
\TT = \widetilde\T^{a^1}\widetilde\T^{a^{2n}}\widetilde\T^{a^{2n-1}}\cdots \widetilde\T^{a^2}: 
\widetilde W^1 \longrightarrow \widetilde W^1
\end{equation}
Our aim will be to check the hypotheses of theorem~\ref{thm-smith} for this map $\TT$.\\

\subsubsection{Partition of walls}
Let us now consider how $\widetilde W^{i}$ may be partitioned according to the sign pattern of $\phi^i-x$.
Following~(\ref{eq-sgnphisi-1}), the coordinate $s_{i-1}$ of this sign is fixed by construction. As for the
coordinate $s_i$, it has been suppressed from $\widetilde W^{i}$ by construction. On the other hand,
proposition~\ref{rem-phisi} implies that this sign is not constant for any other coordinate $j$. Then, denoting
\[
J_i = \NN{n}\setminus\{s_{i-1},s_i\},
\]
it follows that the above mentioned regions may be described using sign vectors in the finite set
$\Sigma_i=\{\pm 1\}^{J_i}$. Vectors in $\Sigma_i$ will then have their coordinates indexed by $J_i$,
which will prove more convenient than indexing them by $\NN{n-2}$. For any sign pattern
$\sigma\in\Sigma_i$, let us define a corresponding 'zone'
\[
{\cal Z}^i(\sigma)=\{x\in \widetilde W^i\,|\,\forall j\in J_i\,,\; \sgn(\phi^i_j-x_j) = \sigma_j\}.
\]
Now, from the translation of all thresholds to zero, the origin of $\R^{n-1}$ belongs to every wall
$\widetilde W^i$. Moreover, no focal point have a coordinate equal to zero. Hence, on each  wall
$\widetilde W^i$, there is a unique zone containing the origin on its boundary, and the corresponding
sign vector in $\Sigma_i$ is
\begin{equation}\label{eq-sigmaij}
\sigma^i = \left(\sgn(\phi^i_j)\right)_{j\in J_i}
\end{equation}
We will see now that these particular zones, having the origin on their boundary, in fact attract all
trajectories of the system, and then remain invariant. As they will be the only zones we consider eventually,
we simply denote:
\[
{\cal Z}^i={\cal Z}^i(\sigma^i).
\]
Remark that the definition of these zones and equation~(\ref{eq-Wi}) allow for a more explicit formulation:
\begin{equation}\label{eq-Zi}
{\cal Z}^i=
\left\{\begin{array}{ll}
	\displaystyle \prod_{j<s_i}[\phi_j^-,\,0]\times\prod_{j>s_i}[0,\,\phi_j^+] & 
	\text{for}\;i\in\NN{n},\\[4mm]
	\displaystyle \prod_{j<s_i}[0,\,\phi_j^+]\times\prod_{j>s_i}[\phi_j^-,\,0] & 
	\text{for}\;i\in\{n+1\cdots 2n\}.
\end{array}\right.
\end{equation}

\subsubsection{Attractive and invariant regions}

\begin{prop}\label{prop-Z1}
The zone ${\cal Z}^1$ is attractive for the dynamics induced by $\TT$:
\[
\TT \left( \widetilde W^1 \right) \subset{\cal Z}^1.
\]
\end{prop}
\begin{proof}
From~(\ref{eq-C}) and~(\ref{eq-sipm}) it follows that $\sgn(\phi_1)=(-1,1\cdots 1)\in\{\pm 1\}^n$.
Then, $s_1=1$ and $s_0\simeq s_{2n}=n$ implies $J_1=\{2\cdots n-1\}$. Then, $\sigma^1 = (1\cdots
1)\in\{\pm 1\}^{J_1}$, and 
\begin{eqnarray*}
{\cal Z}^1 & = & \{x\in \widetilde W^1\,|\,\forall j\in J_1\,,\; x_j <\phi_j^1 \} \\
         & = & \prod_{j=2}^{n}[0,\,\phi_j^+], \qquad\text{since $\phi_j^1=\phi_j^+$ for $j\in J_1$.}
\end{eqnarray*}
Let $x^1\in \widetilde W^1$. In the following we denote $x^i = \widetilde\T^{a^{i}}x^{i-1}$ for
$i\geqslant 2$, so that $\TT x^1 = x^{2n+1}$. Clearly, $x^i\in \widetilde W^i$ for all $i$.\\
Now, for each $i\in\NN{2n}$, the alignment condition~(\ref{eq-alignloop}) and the expression~(\ref{eq-Tai})
of $\T^{a^i}$ imply the following conservation of signs: 
\[
\forall i\in\NN{2n},\,\forall j\ne s_{i+1},\quad \sgn(\phi^i_{j} - x^i_{j}) = 
\sgn(\phi^{i+1}_{j} - x^{i+1}_{j}).
\]
In particular for the $n$ last domains of the cycle $\cal C$, remarking from table~(\ref{eq-sipm}) that
$s_{n+i}=i$, the above has the following consequence:
\[
\forall i\in\NN{n},\,\forall j < i+1,\quad \sgn(\phi^{n+i}_{j} - x^{n+i}_{j}) = 
\sgn(\phi^{n+i+1}_{j} - x^{n+i+1}_{j}).
\]
This equality can be propagated, in the sense that 
\[
\begin{array}{lcl}
\sgn(\phi^{n+1}_{1} - x^{n+1}_{1}) & = & \sgn(\phi^{n+2}_{1} - x^{n+2}_{1})
\,=\, \dots \,=\, \sgn(\phi^{2n}_{1} - x^{2n}_{1}) \,=\, \sgn(\phi^{1}_{1} - x^{2n+1}_{1}), \\[2mm]
\sgn(\phi^{n+2}_{2} - x^{n+2}_{2}) & = & \sgn(\phi^{n+3}_{2} - x^{n+3}_{2})
\,=\, \dots \,=\, \sgn(\phi^{2n}_{2} - x^{2n}_{2}) \,=\, \sgn(\phi^{1}_{2} - x^{2n+1}_{2}),\\
 & \vdots & \\
\sgn(\phi^{2n-1}_{n-1} - x^{2n-1}_{n-1}) & = & \sgn(\phi^{2n}_{n-1} - x^{2n}_{n-1}) 
\,=\, \sgn(\phi^{1}_{n-1} - x^{2n+1}_{n-1}) \\[2mm]
\sgn(\phi^{2n}_{n} - x^{2n}_{n}) & = & \sgn(\phi^{1}_{n} - x^{2n+1}_{n}) 
\end{array}
\]
Now, since $x^{n+i}\in\widetilde W^{n+i}$, and since~(\ref{eq-sgnphisi-1}) is satisfied on $\widetilde
W^{n+i}$ by definition, one obtains
\begin{equation}\label{eq-conservsgn}
\sgn(\phi^{n+i+1}_{i} - x^{n+i+1}_{i}) = \sgn(\phi^{n+i+1}_{i}),\qquad i\in\NN{n}.
\end{equation}
Combining this with the previous list of equalities gives 
\[
\sgn(\phi^{1}_{i} - x^{2n+1}_{i}) = \sgn(\phi^{1}_{i}),\qquad i\in\NN{n},
\]
which exactly means $x^{2n+1}=\TT x^{1}\in{\cal Z}^1$, as expected.
\end{proof}

One may now consider the restriction of $\TT$ to ${\cal Z}^1$. 
Actually, the previous proposition implies that chosing any intial condition $x^1\in\widetilde W^1$, the
trajectory $\{\TT^m x\,|\, m\in\N\}$ is contained in ${\cal Z}^1$, except maybe for $x^1$.
So, there is no loss of generality in chosing initial conditions in this invariant subset only.\\
We can also verify that, on each wall all trajectories of system~(\ref{eq-sysloopp}) are eventually
contained in the region ${\cal Z}^i$. This fact is a consequence of
proposition~\ref{prop-Z1} and the following:
\begin{prop}\label{prop-Zi}
For all $i\in\NN{2n}$, 
$\quad\widetilde\T^{a^{i}}\left({\cal Z}^{i-1} \right) \subset{\cal Z}^{i}$.
\end{prop}
\begin{proof}
Let $x\in {\cal Z}^i$, which by definition is equivalent to
\begin{equation}\label{eq-sgnZi}
\forall j\in \NN{n}\setminus\{s_{i-1}\},\qquad \sgn(\phi_j^i - x_j) = \sgn(\phi_j^i). 
\end{equation}
First, observe that in dimension $n=2$, $J_i=\varnothing$ by definition and thus ${\cal
Z}_i(\sigma^i)=\widetilde W^{i}$. There is nothing to prove in this case, and one thus assumes
$n\geqslant 3$ in the rest of the proof.\\
From the alignment condition~(\ref{eq-alignloop}) one gets,
\[
\forall j\ne s_{i},\qquad \phi^{i}_j - \widetilde\T^{a^{i}}_jx = (\phi_j^{i-1}-x_j)\,\alpha_j(x).
\]
Since $\alpha_j(x)>0$, using~(\ref{eq-sgnZi}) leads to
\[
\forall j\in\NN{n}\setminus\{s_{i-1},s_i\},\qquad 
\sgn(\phi^{i}_j-\widetilde\T^{a^{i}}_jx) = \sgn(\phi_j^{i-1}) = \sgn(\phi_j^{i}).
\]
From~(\ref{eq-sgnphisi-1}) it is immediate that
\[
\sgn(\phi^{i}_{s_{i-1}}-\widetilde\T^{a^{i}}_{s_{i-1}}x) = \sgn(\phi_{s_{i-1}}^{i-1}) = 
\sgn(\phi_{s_{i-1}}^{i}),
\]
which terminates the proof.
\end{proof}

\begin{figure}
\begin{center}
\scalebox{0.9}{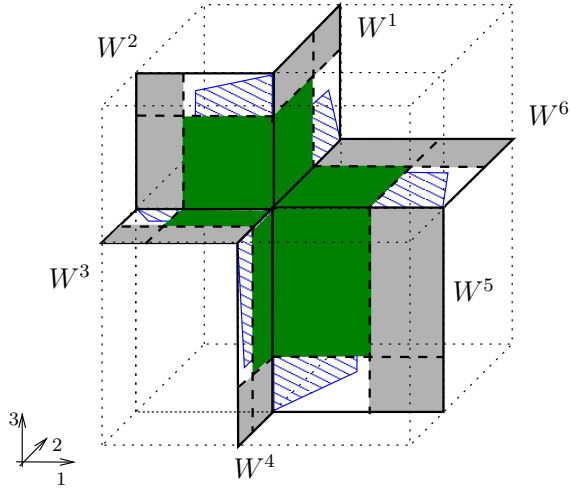}
\caption{\label{fig-invregion}The same walls as in figure~\ref{fig-firstcut}. Now, the invariant region
$\bigcup_i{\cal Z}^i$ has been shaded in dark green. The reader may check on the figure that trajectories
originating on any wall eventually fall into this region.}
\end{center}
\end{figure}

This proposition allows us to consider only the restrictions:
\[
\widetilde\T^{a^i}:{\cal Z}^{i-1}\to {\cal Z}^{i}
\]
in the following. Figure~\ref{fig-invregion} shows the domains ${\cal Z}^i$ on a $3$ dimensional
example.\\

\subsection{Main result}
The main result of this paper will essentially consist in applying theorem~\ref{thm-smith} to feedback loop
systems of the form~(\ref{eq-sysloop}), or equivalently~(\ref{eq-sysloopp}). Among the hypotheses to be
checked, the concavity condition $(C)$ will be verified with the aid of the following lemma.

\begin{lemma}\label{lem-C}
Let $p,q,r\in\mathring{\R}_+^n$. Let also $T:[0,p]\to[0,q]$ and $M:[0,q]\to[0,r]$ be twice
differentiable  mappings satisfying condition $(M)$ of theorem~\ref{thm-smith}. Suppose that any of
the second order derivatives of $T$ and $M$ are nonpositive, and that at least one of them is
negative, for both mappings.\\
Then, their composite $MT:[0,p]\to[0,r]$ satisfies $(M)$ and $(C)$.
\end{lemma}
\begin{proof}
The result is obtained from the chain rule. Let $x\in[0,p]$. Then $Tx\in[0,q]$ and
\[
D\left(MT\right)(x)= DM\left(Tx\right)DT(x).
\]
From this expression it is clear that the composite $MT$ satisfies $(M)$ if both $T$ and $M$ do.  Now,
given given any $x<y$ in $[0,p]$, condition $(M)$ on $M$ implies $Tx < Ty$, by monotonicity of each
coordinate function $T_i$, $M_i$.\\
Then the nonpositivity of second order derivatives gives $DM(Ty) \lneq DM(Tx)$. Then using both 
conditions $(M)$ and $(C)$ on $T$, and condition $(M)$ on $M$, the chain rule expression leads to
\[
D\left(MT\right)(y)= DM\left(Ty\right)DT(y)\lneq DM\left(Ty\right)DT(x)\lneq DM\left(Tx\right)DT(x) = D\left(MT\right)(x).
\]
\end{proof}

We are now in good position to state our main result.

\begin{theo}\label{thm-main}
Consider the map $\TT : {\cal Z}^1 \to {\cal Z}^1$ defined in the previous section.\\
\begin{itemize}
\item If $n=2$, then $\,\forall x \in {\cal Z}^1,\; \TT^m x\to 0$ when $m\to\infty$.\\
\item If $n>2$, then there exists a unique nonzero fixed point $q=\TT q$. Moreover,  $q\in \mathring{\cal
Z}^1$ and for every $x\in{\cal Z}^1\setminus\{0\}$, $\,\TT^m x\to q$ as $m\to\infty$.\\
\end{itemize}
\end{theo}

\begin{proof}
The proof consists in checking the hypotheses of theorem~\ref{thm-smith}, which then yields
precisely the above proposition.\\
First, the upper corner of ${\cal Z}^1$ is the point $(\phi_2^+ \dots \phi_{n}^+)$, which belongs to
$\mathring\R^{n-1}_+$. 
Second, it is clear from the expression~(\ref{eq-Tai}) that each $\widetilde\T^{a^i}$ is $C^1$.
Hence, $\TT$ is $C^1$ one the whole rectangle ${\cal Z}_1$. It follows in particular that $D\TT(0)$ is
well defined.\\
Since the proof is a little lengthy, it will be structured in three parts.\\

\noindent
{\bf Preliminary change of variables.}\\
Before checking the hypotheses of theorem~\ref{thm-smith}, it will prove convenient to apply a local transformation
$\rho^i:{\cal Z}^i\to{\cal P}^i$ at each wall, such that ${\cal P}^i\subset\R_+^{n-1}$. This will allow us to
deal with nonnegative variables only. \\
According to~(\ref{eq-sigmaij}), $\sigma^i_j$ denotes the sign of $\phi_j^i$, for $j\in J_i$.
From~(\ref{eq-sgnZi}), $\phi^i_j$ and $x^i_j$ have the same sign, for $x^i\in{\cal Z}^i$. We also note
$\sigma^i_{j}$ the sign of $\phi_j^i$ when $j\in\{s_{i-1},s_i\}$.  Then, simply define the $j$th coordinate
of this tranformation as
\begin{equation}\label{eq-rhoij}
\rho_j^{i}(x)= \sigma^i_j\:x_j.
\end{equation}
In other words, $\rho^i$ is just a multiplication by a diagonal matrix, whose $j$th entry is
$\sigma^i_j\in\{\pm 1\}$. This implies that $\rho^i$ is invertible, and equals its inverse. Hence we also
denote $\rho^i$ this inverse: $\rho^i:{\cal P}^i\to{\cal Z}^i$. One may also formulate ${\cal P}^i$
explicitly:
\begin{equation}\label{eq-Pi}
{\cal P}^i=\prod_{j\ne s_i} \left[0,|\phi_j^i|\,\right].
\end{equation}

It is tempting now to define maps $\M^{(i)}$, by the condition that the following diagram commutes, for any
$i\in\NN{2n}$:
\begin{equation}\label{eq-Mi}
\begin{array}{ccc}
{\cal Z}^{i-1}                & \xrightarrow{\T^{a^i}} & {\cal Z}^{i} \\
\Big\downarrow \stackrel{\rho^{i-1}}{} &              & \Big \downarrow \stackrel{\rho^{i}}{}\\
{\cal P}^{i-1}     & \xrightarrow{\M^{(i)}} & {\cal P}^i
\end{array}
\end{equation}
With the previous remark on $\rho^i$ this may also be written as ${\M}^{(i)}=\rho^i\: 
\T^{a^i}\rho^{i-1}$. Composing the diagrams above for successive $i$, and since $\rho^1=\rho^{2n}=id$, one
gets ${\cal P}^1={\cal Z}^1$, and
\begin{equation}\label{eq-TTM}
\TT= {\M}^{(1)}{\M}^{(2n)}{\M}^{(2n-1)}\cdots {\M}^{(2)}
\end{equation}
For any $x^1\in{\cal P}^1=[0,p]$ we write $x^i = \widetilde{\M}^{(i)}x^{i-1},$ $i\geqslant 2$. From
proposition~\ref{prop-Zi}, one gets $x^i\in{\cal Z}^i$.\\
Each ${\M}^{(i)}$ can be expressed coordinate-wise:
\[
{\M}^{(i)}_j x = \sigma^i_j\phi^i_j + \sigma^i_j\:\left( \sigma^{i-1}_j\:x_j - \phi^i_j\right)
\left(\frac{\phi^i_{s_i}}{\phi^i_{s_i}-\sigma^{i-1}_{s_i}\:x_{s_i}}\right)^
{\frac{\gamma_j}{\gamma_{s_i}}} \qquad j\ne s_i.
\]
From their definition, the $\sigma^i_j$s clearly satisfy the alignment condition~(\ref{eq-alignloop}):
$\sigma^i_j=\sigma^{i-1}_j$ for $j\ne s_i$. It also comes from~(\ref{eq-alignloopbis}) that:
$
\sigma^i_{s_i}= -\sigma^{i-1}_{s_i}
$. 
Hence, one also has:
\begin{equation}\label{eq-Mij}
{\M}^{(i)}_j x = |\phi^i_j|+\left( x_j - |\phi^i_j|\right)\alpha_j\left(\sigma^{i-1}_{s_i}\:x_{s_i}\right) 
\qquad j\ne s_i,
\end{equation}
where
\[
\alpha_j\left(\sigma^{i-1}_{s_i}\:x_{s_i}\right) 
= \left(\frac{\phi^i_{s_i}}{\phi^i_{s_i}-\sigma^{i-1}_{s_i}\:x_{s_i}}\right)^
  {\frac{\gamma_j}{\gamma_{s_i}}}
= \left(\frac{|\phi^i_{s_i}|}{|\phi^i_{s_i}|+ x_{s_i}}\right)^ {\frac{\gamma_j}{\gamma_{s_i}}}
\]

\noindent
{\bf Monotonicity and concavity.}\\
One computes now:
\begin{equation}\label{eq-dMij}
\frac{\partial\M^{(i)}_j}{\partial x_k} (x) =\left\{
\begin{array}{ll}
\displaystyle \alpha_j\left(\sigma^{i-1}_{s_i}\:x_{s_i}\right) & \text{if }\;  j=k\\[2mm]
\displaystyle\frac{\gamma_j}{\gamma_{s_i}}\,\frac{|\phi^i_j| - x_j}{|\phi^i_{s_i}|+ x_{s_i}}\,
\alpha_j\left(\sigma^{i-1}_{s_i}\:x_{s_i}\right) &
\text{if }\;  k=s_i\\[2mm]
0 & \text{otherwise}.
\end{array}\right.
\end{equation}
and
\begin{equation}\label{eq-d2Mij}
\frac{\partial^2\M^{(i)}_j}{\partial x_m\partial x_k} (x) =\left\{
\begin{array}{ll}
\displaystyle -\frac{\gamma_j}{\gamma_{s_i}} \,\frac{\alpha_j\left(\sigma^{i-1}_{s_i}\:x_{s_i}\right)}
{|\phi^i_{s_i}|+ x_{s_i}}
& \begin{array}{l} \text{if }\;  j=k,\,m=s_i\\[1mm] \text{or } k=s_i,\,m=j\end{array} \\[4mm]
\displaystyle- \frac{\gamma_j}{\gamma_{s_i}} \left(1+\frac{\gamma_j}{\gamma_{s_i}}\right) 
\frac{|\phi^i_j|- x_j}{(|\phi^i_{s_i}|+ x_{s_i})^2} \,\alpha_j\left(\sigma^{i-1}_{s_i}\:x_{s_i}\right) & 
\text{if }\; m=k=s_i\\[3mm]
0 & \text{otherwise}.\\
\end{array}\right.
\end{equation}
One sees that ${\M}^{(i)}$ is very similar to $\T^{a^i}$, so that equations above are strongly reminiscent
of~(\ref{eq-dTidxj}) and~(\ref{eq-d2Tidxj}). Let us evaluate the sign of these quantities. From $x\in{\cal
P}^{i-1}$ and~(\ref{eq-Pi}) it follows that $|\phi^i_j|-x_j\geqslant 0$. It is clear also that 
$|\phi^i_{s_i}|+ x_{s_i}>0$. It follows that, for all maps $\M^{(i)}$, the nonzero terms of the jacobian 
are positive, while the nonzero derivatives of order $2$ are negative, at any point $x\in{\cal P}^{i-1}$.\\
Applied to~(\ref{eq-TTM}), the chain rule gives
\begin{equation}\label{eq-chain}
D\TT(x^1) =D{\M}^{(1)}(x^{2n}) D{\M}^{(2n)}(x^{2n-1}) \cdots D{\M}^{(2)}(x^1).
\end{equation}
Moreover, it seems from~(\ref{eq-dMij}) that the jacobians in the product above are the sum of a diagonal
matrix and a matrix whose column number $s_i$ contains the only nonzero elements.\\
Some care must be taken though, due to the particular indexing of matrices. Actually, from~(\ref{eq-dMij}) the
diagonal entries of $D{\M}^{(i)}$ are at first sight $D{\M}^{(i)}_{jj} = \alpha_j
\left(\sigma^{i-1}_{s_i}\:x^{i-1}_{s_i}\right)$. However, the row subscripts of $D{\M}^{(i)}$ belong to
$\NN{n}\setminus\{s_i\}$, while the columns are indexed by $\NN{n}\setminus\{s_{i-1}\}$. This implies that
neither $D{\M}^{(i)}_{s_i,s_i}$ nor $D{\M}^{(i)}_{s_{i-1},s_{i-1}}$ are defined. The exact shape of
nonzero entries in $D{\M}^{(i)}$ depends on the order in which $s_{i-1}$ and $s_{i}$ are sorted. In our
case, table~(\ref{eq-sipm}) shows that either $s_i=s_{i-1}+1$, or $s_{i}=1$ and $s_{i-1}=n$, the latter
occurring for $i\in\{1,n+1\}$.\\
Let us represent the subscripts of nonzero entries in both cases. First, if $i\not\in\{1,n+1\}$, both 
$D{\M}^{(i)}$ and $D{\M}^{(n+i)}$ have the following shape:
\begin{equation}\label{eq-shapeDM}
{\scriptsize
\begin{array}{l}
\begin{array}{ccccccc}
\phantom{(1,1) } & \phantom{\ddots} & \phantom{(s_{i-1}-1,s_{i-1}-1) } & 
\phantom{xxxx}{\stackrel{\displaystyle s_{i-1}}{\displaystyle \downarrow}} & & & \\
\end{array}\\[4mm]
\left[\begin{array}{ccccccc}
(1,1) &   &                       & (1,s_i)         &                &  & \\
 & \ddots &                       & \vdots          &                &  & \\
 &        & (s_{i-1}-1,s_{i-1}-1) & (s_{i-1}-1,s_i) &                &  & \\
 &        &                       & (s_{i-1},s_i)   &                &  & \\
 &        &                       & (s_i+1,s_i)     & (s_i+1,s_i+1)  &  & \\
 &        &                       & \vdots          &                & \ddots &     \\
 &        &                       & (n,s_i)         &                &        & (n,n) 
\end{array}\right]
\end{array} }
\end{equation}
Now, if $i\in\{1,n+1\}$ which implies that row $1$ and column $n$ are removed:
\begin{equation}\label{eq-shapeDMbis}
D{\M}^{(i)}\equiv {\scriptsize
\left[\begin{array}{cccc}
(2,1)   & (2,2) &        &           \\
\vdots  &       & \ddots &           \\
\vdots  &       &        & (n-1,n-1) \\
(n,1)   & 0     & \cdots &  0
\end{array}\right]}
\end{equation}
Since $s_{i-1}$ ranges over $\NN{n-1}$ when $i$ or $n+i$ varies in $\{2\cdots n\}$, it is not hard
to deduce from~(\ref{eq-shapeDM}) that both products:
\[
\Pi^2=D{\M}^{(2n)}(x^{2n-1}) \cdots D{\M}^{(n+2)}(x^{n+1})\qquad\text{and}\qquad
\Pi^1={\M}^{(n)}(x^{n-1}) \cdots D{\M}^{(2)}(x^1)
\]
have no zero entries, and are thus positive. Then, from~(\ref{eq-shapeDMbis}) one deduces that
\[
D\TT(x^1) = D{\M}^{(1)}(x^{2n})\,\Pi^2\, D{\M}^{(n+1)}(x^n)\,\Pi^1
\]
is positive as well. Hence we have just shown $(M)$:
\[
\forall x^1\in{\cal P}^1,\quad D\TT(x^1) > 0.
\]

As for concavity, it follows from the negative terms in~(\ref{eq-d2Mij}) and lemma~\ref{lem-C} that 
\[
D\left({\M}^{(i+1)}{\M}^{(i)}\right)(x^{i-1})\lneq D\left({\M}^{(i+1)}{\M}^{(i)}\right)(y^{i-1}).
\]
Then a simple induction shows that $D\TT$, as expressed in~(\ref{eq-chain}), satisfies condition $(C)$:
\[
\forall x^1,y^1\in{\cal P}^1,\;0<x^1<y^1,\quad D\TT(y^1) \lneq D\TT(x^1).
\]

\noindent
{\bf Behaviour at the upper corner and the origin.}\\
Before studying the origin, let us show that $\TT p<p$, as required by theorem~\ref{thm-smith}. Here,
$p=(\phi_2^+\dots \phi_{n}^+)$ is the upper corner of ${\cal Z}^1$, \ie ${\cal Z}^1=[0,p]$.  Now, let us
consider the last maps composing $\TT$, as in the proof of proposition~\ref{prop-Z1}. Namely, one deduces
from~(\ref{eq-Taisi-1}) that:
\[
\forall x\in\widetilde W^{n+i-1},\qquad
\widetilde\T^{a^{n+i}}_{i-1}x = \left(1- \alpha_{i-1}(x)\right)\phi^{n+i}_{i-1},
\]
As usual, one identifies $a^{2n+1}$ and $a^1$. Then for $i\in\{2\cdots n+1\}$,
$\alpha_{i-1}(x)\in(0,1]$ gives:
\[
\widetilde\T^{a^{n+i}}_{i-1}x = \left(1- \alpha_{i-1}(x)\right)\phi^{+}_{i-1} < \phi^{+}_{i-1}.
\]
Now, from equation~(\ref{eq-conservsgn}) in the proof of proposition~\ref{prop-Z1}, one knows that
$\sgn(\phi^{n+i}_{s_{i-1}}-x_{s_{i-1}})$ is conserved through the last steps of the cycle, so that
\[
\forall i\in\{2\cdots n\},\,\forall x\in\widetilde W^{n+i-1},\qquad
\left(\widetilde\T^{a^{1}} \widetilde \T^{a^{2n}} \cdots \widetilde\T^{a^{n+2}}\right)_i x < \phi_i^+
\]
Since in particular 
\[
\widetilde\T^{a^{n+1}} \widetilde \T^{a^{n}} \cdots \widetilde\T^{a^{2}}p\in\widetilde W^{n+1},
\]
it follows that $\TT p<p$.\\

To terminate the proof, it remains to check whether the spectral radius of $D\TT(0)$ is greater than $1$ or
not. First, remark that $\TT(0)=0$ implies that each $D{\M}^{(i)}$ has to be evaluated at $0$ in
expression~(\ref{eq-chain}). Then from the expression of $\alpha_j$ in~(\ref{eq-dMij}), for $i\in\{2\cdots
n\}$ the diagonal terms of each $D{\M}^{(i)}(0)$ equal $1$, except at $(i-1,i-1)$. Here, it is given by the
second line in~(\ref{eq-dMij}), just as the rest of column $i-1$. The latter is given by:
\[
\forall i\in\{2\cdots n\}\qquad
 D{\M}^{(i)}_{j,\,i-1}(0) = \frac{\gamma_j \,|\phi^i_j|}{\gamma_{i}\, |\phi^i_{i}|}
= \left| \frac{\kk^i_j}{\kk^i_i}\right|,
\]
with the notation $\kk^i_j=\kk_j(a^i)$.\\
If the case $n=2$, matrices $D{\M}^{(i)}(0)$ are in fact scalars of the form above. From
equations~(\ref{eq-sysloop}) and~(\ref{eq-sysloopp}) one obtains 
\[
\kk^1_2=\kk^4_2=\kk^+_2,\quad \kk^4_1=\kk^3_1=\kk^+_1,\quad  
\kk^3_2=\kk^2_2=\kk^-_2,\quad\text{ and }\quad\kk^1_1=\kk^2_1=\kk^-_1.
\]
See figure~\ref{fig-plan} for the state space of two dimensional feedback loops.
Then,~(\ref{eq-chain}) gives
\[
D\TT(0)=\left|\frac{\kk^1_2}{\kk^1_1}\,\frac{\kk^4_1}{\kk^4_2}\,\frac{\kk^3_2}{\kk^3_1}
\,\frac{\kk^2_1}{\kk^2_2}\right| = 1.
\]

\begin{center}
\begin{figure}
\begin{center}
\scalebox{0.6}{\input{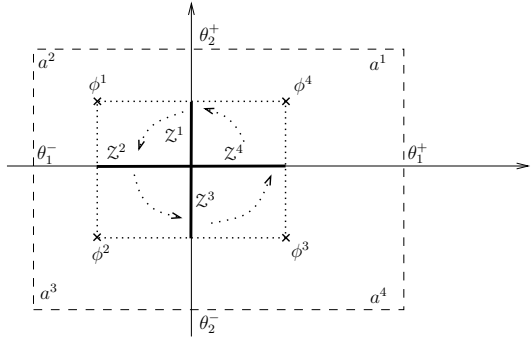}}
\caption{\label{fig-plan}The state space of a feedback loop with only two variables, schematically. The
notations are those explained in the text.}
\end{center}
\end{figure}
\end{center}
Suppose now that $n\geqslant 3$. Before going further, let us remark the following fact: if a matrix
$A$ is such that $A-Id >0$, then the spectral radius $\rho(A)>1$. This is a direct consequence of
Perron-Frobenius theorem. Actually, $\rho(A)$ is the maximal eigenvalue of $A$, or the maximal root of
its characteristic polynomial, which we denote $\pi_A(X)$. One clearly has $\pi_{A-Id}(X)=\pi_A(X-1)$,
so that the roots of $\pi_{A-Id}$ are exactly those of $\pi_A$ minus one. Applying Perron-Frobenius 
theorem to the positive matrix  $A-Id$ implies $\rho(A)>1$.\\
Hence to terminate the proof, it suffices to show that $D\TT(0)-Id>0$, or equivalently that the
diagonal terms of $D\TT(0)$ are strictly greater than $1$, since we have shown $(M)$ already. We
use~(\ref{eq-chain}), from which it is possible to derive the general term of $D\TT$.\\
The subscripts in~(\ref{eq-shapeDM}) and~(\ref{eq-shapeDMbis}) are those to be used in
equation~(\ref{eq-dMij}). To compute~(\ref{eq-chain}) on the other hand, it is more relevant to use
the actual row and column numbers, as represented in~(\ref{eq-shapeDM}) over the nonzero column, which
is numbered $s_{i-1}$. Hence in the expressions below, $D\M^{(i)}_{jk}$ refers to the entry $j,k$ of
matrix $D\M^{(i)}(0)$, which is not necessarily $\displaystyle\frac{\partial\M^{(i)}_j}{\partial
x_k}(0)$.
A simple induction shows that
\[
\left(D\TT(x^1)\right)_{ij} = \sum_{k_1,\dots k_{2n-1}=1}^{n-1} D\M^{(1)}_{ik_1}D\M^{(2n)}_{k_1k_2}
                              \cdots D\M^{(2)}_{k_{2n-1}j} 
\]
In particular, the diagonal terms can be conveniently denoted using the following abbreviation. Define the
integer vector ${\bf k}=(k_1\dots k_{2n-1})\in\NN{n-1}^{2n-1}$, and denote
\[
p_{\bf k}^i = D\M^{(1)}_{ik_1}D\M^{(2n)}_{k_1k_2}\cdots D\M^{(2)}_{k_{2n-1}i}.
\]
Note that the general form of terms in products above is the following:
\[
 D\M^{(j)}_{k_{2n+1-j}k_{2n+2-j}}\qquad\quad j\in\{3\cdots 2n\}.
\]
Then, $\left(D\TT(x^1)\right)_{ii}$ equals the sum of products $p_{\bf k}^i$ when $\bf k$ varies
in the whole set $\NN{n-1}^{2n-1}$.\\
Now, since each matrix $D{\M}^{(i)}(0)$ is composed of zeros and positive terms, it follows that each
nonzero product $p_{\bf k}^i$ is positive. Hence, if for all $i$ there is one $\bf k$ such that $p_{\bf
k}^i=1$, and at least one other nonzero $p_{\bf k}^i$, the proof will be finished.\\

To help intuition let us detail the shape of $D{\M}^{(i)}(0)$ and $D{\M}^{(n+i)}(0)$ for $n\geqslant i>1$,
thanks to~(\ref{eq-shapeDM}) and~(\ref{eq-shapeDMbis}):
\begin{equation}\label{eq-DM0}
{\scriptsize
\begin{array}{l}
\begin{array}{ccccccc}
\phantom{1 }  & \phantom{\ddots}& \phantom{1} & 
\phantom{xxxx}{\stackrel{\displaystyle i-1}{\displaystyle \downarrow}} & & & \\
\end{array}\\[4mm]
\left[\begin{array}{ccccccc}
1 &       &   & |\kk^i_1/\kk^i_{i}|         &   &  & \\[1mm]
 & \ddots &   & \vdots          &   &  & \\[1mm]
 &        & 1 & |\kk^i_{i-2}/\kk^i_{i}| &   &  & \\[1mm]
 &        &   & |\kk^i_{i-1}/\kk^i_{i}|   &   &  & \\[1mm]
 &        &   & |\kk^i_{i+1}/\kk^i_{i}|     & 1 &  & \\[1mm]
 &        &   & \vdots          &   & \ddots &     \\[1mm]
 &        &   & |\kk^i_n/\kk^i_{i}|         &   &        & 1 
\end{array}\right]
\end{array} }
\end{equation}
For $i\in\{1,n+1\}$:
\begin{equation}\label{eq-DM0bis}
D{\M}^{(i)}\equiv {\scriptsize
\left[\begin{array}{cccc}
|\kk^i_2/\kk^i_1|  & 1 &        &           \\
\vdots  &   & \ddots &           \\
\vdots  &   &        & 1 \\
|\kk^i_n/\kk^i_1|  & 0 & \cdots &  0
\end{array}\right]}
\end{equation}
Now, we claim that $p_{\bf k}^i=1$, for $1<i\leqslant n-1$, and 
\begin{equation}\label{eq-k}
{\bf k}=(1\cdots 1,i\cdots i,1\cdots 1,i\cdots i)
\end{equation}
where $k_j=i$ for $n+1-i\leqslant j\leqslant n$, and $2n+1-i\leqslant j\leqslant 2n-1$, and $k_j=1$ otherwise. 
Actually, one has on the one hand:
\[
D\M^{(1)}_{i,k_1}=D\M^{(1)}_{i,1}=\left|\frac{\kk^1_{i+1}}{\kk^1_1}\right| \qquad\text{and}\qquad
D\M^{(n+1)}_{k_n,k_{n+1}}=D\M^{(n+1)}_{i,1}=\left|\frac{\kk^{n+1}_{i+1}}{\kk^{n+1}_1}\right|
\]
and on the other hand
\[
D\M^{(n+i+1)}_{k_{n-i},k_{n+1-i}}=D\M^{(n+i+1)}_{1,i}=\left|\frac{\kk^{n+i+1}_1}{\kk^{n+i+1}_{i+1}}\right|
\qquad\text{and}\qquad
D\M^{(i+1)}_{k_{2n-i},k_{2n+1-i}}=D\M^{(i+1)}_{1,i}=\left|\frac{\kk^{i+1}_1}{\kk^{i+1}_{i+1}}\right|.
\]
Moreover, for any $j\in\{2n+1-i\cdots 2n-1\}\cup\{n+1-i\cdots n\}$, it appears that $D\M^{(j)}_{i,i}=1$,
since $i\ne s_{j}-1$, and similarly, for $j\in\{1\cdots n-i\}\cup\{n+1\cdots 2n-i\}$ one has 
$D\M^{(j)}_{1,1}=1$. As a consequence, for $\bf k$ as in~(\ref{eq-k}), 
\begin{eqnarray}\label{eq-pki}
p_{\bf k}^i &=& D\M^{(1)}_{i,1}\,D\M^{(n+i+1)}_{1,i}\,D\M^{(n+1)}_{i,1}\,D\M^{(i+1)}_{1,i}\nonumber\\ 
 &=& \left|\frac{\kk^1_{i+1}}{\kk^1_1}\frac{\kk^{n+i+1}_1}{\kk^{n+i+1}_{i+1}}
           \frac{\kk^{n+1}_{i+1}}{\kk^{n+1}_1}\frac{\kk^{i+1}_1}{\kk^{i+1}_{i+1}}\right|
\end{eqnarray}
Now we refer to the table~(\ref{eq-C}) of successive boxes, and recall that $\phi^i$ belongs to the domain
$a^{i+1}$. Then it is rather straightforward to check that
\[
\sgn(\phi^{n+i})=-\sgn(\phi^i)
\]
for any $i\in\NN{n}$. It follows immediately that $\sgn(\kk^{n+i})=-\sgn(\kk^i)$ as well. From this and the
fact that any production term $\kk^i_j$ is fully determined by its sign, it follows that
\[
\frac{\kk^{i+1}_1\,\kk^{n+i+1}_1}{\kk^1_1\;\kk^{n+1}_1}=1\qquad\text{and}\qquad
\frac{\kk^1_{i+1}\,\kk^{n+1}_{i+1}}{\kk^{i+1}_{i+1}\,\kk^{n+i+1}_{i+1}}=1,
\] 
whence $p_{\bf k}^i= 1$.\\

Recall that the previous holds only for $i>1$. Now, let ${\bf k}=(1\cdots 1)$. Then, 
\[
D\M^{(1)}_{1,k_1}=D\M^{(1)}_{1,1}=\left|\frac{\kk^1_{2}}{\kk^1_1}\right| \qquad\text{and}\qquad
D\M^{(n+1)}_{k_n,k_{n+1}}=D\M^{(n+1)}_{1,1}=\left|\frac{\kk^{n+1}_{2}}{\kk^{n+1}_1}\right|
\]
while
\[
D\M^{(n+2)}_{k_{n-1},k_{n}}=D\M^{(n+2)}_{1,1}=\left|\frac{\kk^{n+2}_1}{\kk^{n+2}_{2}}\right|
\qquad\text{and}\qquad
D\M^{(2)}_{k_{2n-1},1}=D\M^{(2)}_{1,1}=\left|\frac{\kk^{2}_1}{\kk^{2}_{2}}\right|.
\]
Moreover for any other $j$, $D\M^{(j)}_{1,1}=1$, and thus
\begin{eqnarray*}
p_{\bf k}^1 &=& \left|\frac{\kk^1_{2}}{\kk^1_1} \frac{\kk^{n+2}_1}{\kk^{n+2}_{2}}
                    \frac{\kk^{n+1}_{2}}{\kk^{n+1}_1}\frac{\kk^{2}_1}{\kk^{2}_{2}}\right|\\
	    &=& 1,\qquad\qquad\qquad\qquad\text{for the same reason as in~(\ref{eq-pki})}.
\end{eqnarray*}
Thus, we have found, for each $i\in\NN{n-1}$, a $\bf k$ such that $p^i_{\bf k}=1$. Now, for any $i>1$, let
${\bf k}=(1\cdots 1,i\cdots i)$, where $k_j=i$ for $2n+1-i\leqslant j\leqslant 2n-1$. This leads to
\[
p^i_{\bf k}=D\M^{(1)}_{i,1}D\M^{(2n)}_{1,1}\cdots D\M^{(i+1)}_{1,i}D\M^{(i)}_{i,i}\cdots D\M^{(2)}_{1,i},
\]
and from~(\ref{eq-DM0}) and~(\ref{eq-DM0bis}) one checks readily that the above is positive: the entry $(1,1)$
is actually never zero in the first terms, the $i$th column of $D\M^{(i+1)}$ is nonzero, and the diagonal
terms of the last terms in the product are positive as well.\\
For $i=1$, a positive $p^1_{\bf k}$ is found for instance with ${\bf k}=(2\cdots 2)$. Actually, $(2,2)$
entries are never zero in~(\ref{eq-DM0}), and moreover $D\M^{(1)}_{1,2}$ and $D\M^{(2)}_{2,1}$ are also
positive.
\end{proof}

\section{Conclusion}\label{sec-ccl}
The main result of this paper is theorem~\ref{thm-main}, which is a direct application of a theorem of
Smith~\cite{smith}, here refered to as theorem~\ref{thm-smith}. It is in fact very similar to the main theorem of
Snoussi~\cite{snoussi}, since the latter states existence of a limit cycle where theorem~\ref{thm-main} states
existence and unicity, for the class of negative feedback loop systems of the form~(\ref{eq-sysloopp}). The
improvement is entirely obtained from the use of theorem~\ref{thm-smith}. This use is made possible by the
natural occurrence of monotone, concave transformations in the context of piecewise affine gene network models of
the form~(\ref{eq-genenet}). In general, these transformations only appear on particular regions in phase space,
as shown in section~\ref{sec-regions}. Also, they are monotone and concave only up to a symmetry (or say with
respect to the order induced by orthants which are not necessarily positive), which depends on the region on
which they are defined. For negative feedback loop systems however, it is shown in section~\ref{sec-restricdom}
that after a finite number of steps, all trajectories belong to one region, where the first-return dynamics is
monotone and concave. It seems promising then, to study systems of a more general from than a negative feedback
loop, using the same tools as in the present paper. For such systems, one should not concentrate on a single
monotone concave transformation, but on several of them, and take into account the possible transitions between
their domains of definition.


\end{document}